\documentclass[aps,prl,twocolumn,superscriptaddress,showpacs]{revtex4}%
\usepackage{amsmath}
\usepackage{graphicx}%

\begin{document}
\title{Objectively discerning Autler-Townes Splitting \\ from Electromagnetically Induced Transparency}
\author{Petr M.\ Anisimov}
\email{petr@lsu.edu}
\author{Jonathan P.\ Dowling}
\affiliation{Hearne Institute for Theoretical Physics and Department of Physics and Astronomy \\
Louisiana State University, Baton Rouge, LA 70803 }
\author{Barry C.\ Sanders}
\affiliation{Institute for Quantum Information Science, University of Calgary,
  Alberta T2N 1N4, Canada}

\date{\today }

\begin{abstract}
Autler-Townes splitting (ATS) and electromagnetically-induced transparency
(EIT) both yield transparency in an absorption profile, but only EIT
yields strong transparency for a weak pump field due to Fano
interference. Empirically discriminating EIT from ATS is important but so far
has been subjective. 
We introduce an objective method, based on Akaike's information
criterion, to test ATS vs.\ EIT from experimental data and determine which
pertains. We apply our method to a recently reported induced-transparency
experiment in superconducting circuit quantum electrodynamics. 
\end{abstract}

\pacs{42.50.Gy, 42.50.Ct}

\maketitle

Coherent processes in atoms and molecules yield many interesting and practical
phenomena such as coherent population trapping~\cite{Arimondo1996}, lasing
without inversion~\cite{Kocharovskaya1992}, and electromagnetically induced
transparency (EIT)~\cite{Marangos1998}. Pioneering EIT experiments employed
alkali metals due to their simple electronic level structure and long-lived
coherence, but recently coherent processes are investigated in other systems
such as  quantum dots~\cite{Xu2008}, nanoplasmonics~\cite{Liu2009},
superconducting circuits~\cite{Kelly2010},
metamaterials~\cite{Papasimakis2008,Zhang2008}, and
optomechanics~\cite{SMC+11}. EIT is also observed for classical coupled
oscillator, e.g.\ inductively or capacitively coupled electrical resonator
circuits~\cite{a:LambRetherford1951,a:Nussenzveig2002}. EIT systems could
enable new practical applications of coherent processes, but the lack of
time-scale separations characteristic of
alkalis~\cite{FleischhauerMarangos2005} obfuscates signatures of coherent
processes. 
 
Here we focus on EIT, where transparency is induced coherently by a pump field
even if the pump is arbitrarily weak. EIT is crucial for optically-controlled
slowing of light~\cite{Hau1999} and optical storage~\cite{Philips2001} and is
achieved by Fano interference~\cite{Fano1961} between two atomic
transitions. Without Fano interference, EIT is simply Autler-Townes splitting
(ATS), equivalently the ac-Stark effect~\cite{AutlerTownes1955}, corresponding
to a doublet structure in the atomic-absorption profile that requires strong
pumping. In essence, both EIT and ATS lead to the presence of a transparency
window due to electromagnetic (EM) pumping, but the mechanisms are entirely
different. Here we introduce an objective test for use on empirical data to
discern EIT from ATS in any experiment. This test is based on Akaike weights
for the models~\cite{BurnhamAnderson2002} and reveals whether EIT or ATS has
been observed or whether the operating conditions make the data inconclusive. 

Fano's seminal study of two nearly-resonant modes decaying via a common
channel differed from the prevalent normal-mode analyses at the time: he
showed that this shared decay channel yields additional cross-coupling between
modes mediated by the common reservoir, which explained the anomalous
asymmetric lineshape for electrons scattering from Helium~\cite{Fano1961}. In
fact any response that combines multiple modes can have Fano interference,
which can be extremely sharp and highly sensitive to variability in the
system~\cite{Hao2008}.  

\begin{figure}
\includegraphics[width=\columnwidth]{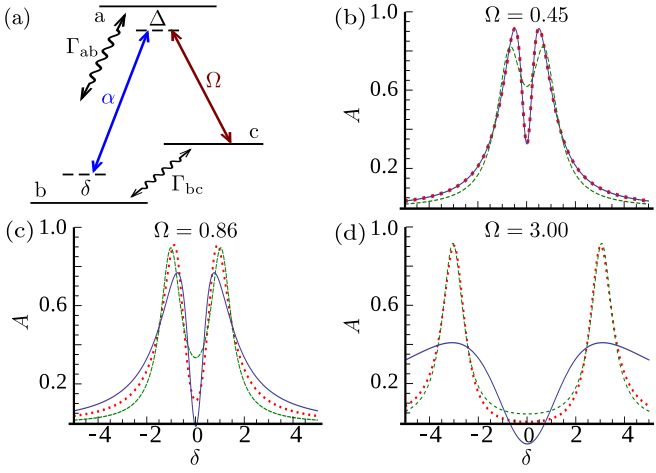}
\caption{(Color online)
	(a)~$\Lambda$-type TLA with probe (pump) driving field with Rabi
	frequency~$\alpha$ ($\Omega$), which probes (drives) the
	$|a\rangle\leftrightarrow|b\rangle$ ($|a\rangle\leftrightarrow|c\rangle$) transition.
	(b-d)~Absorption~$A$ vs.\ two-photon detuning~$\delta$
	(red dots) for resonant ($\Delta=0$) pump with $\Gamma_\text{ab}=1$,
	$\Gamma_\text{bc}=0.1$ and various $\Omega$ with best fits
	to $A_\text{EIT}(C_+,C_-,\gamma_+,\gamma_-)$ (blue solid) and
	$A_\text{ATS}(C,\gamma,\delta_{0})$ (green dashed) models calculated for
	(b)~weak $\Omega$ with good fit to
	$A_\text{EIT}(2.14,1.89,0.581,0.520)$ and poor fit to closest
	$A_\text{ATS}(0.532,0.633,0.712)$, (c)~intermediate $\Omega$ with poor
	fit to closest $A_\text{ATS}(0.472,0.512,1.03)$ as well as
	$A_\text{EIT}(88.3,88.3,0.75,0.752)$, and (d)~strong $\Omega$ with
	poor fit to closest $A_\text{EIT}(1.3\times 10^3,1.3\times
	10^3,2.92,2.92)$ and good fit to $A_\text{ATS}(0.499,0.521,3.05)$. 
	}
\label{fig:LS}
\end{figure}

Harris and Imamo\u{g}lu showed that hybrid ``atom+field'' modes in the
dressed-state formalism interact with the same reservoir hence readily satisfy
the Fano interference conditions~\cite{ImamogluHarris89} thereby producing a
transparency window in the absorption profile~$A(\delta)$ for~$\delta$ the
two-photon detuning frequency. This effect was originally demonstrated for a
$\Lambda$-type three-level atom (TLA) with energy levels~$|a\rangle$,
$|b\rangle$, and $|c\rangle$ and judiciously chosen rates as shown in
Fig.~\ref{fig:LS}(a). Dressed-state frequency separation is proportional to
the pump-field Rabi frequency~$\Omega$, and this separation yields ATS in the
absence of Fano interference. Fano interference is negligible for
large~$\Omega$ but must transition smoothly from ATS to EIT as $\Omega$
decreases and the dressed states try to merge thereby strengthening the Fano
interference effect. Under EIT conditions, complete transparency holds even in
the weak-pump limit. 

There are four TLAs: $\Lambda$, V, and two ladder ($\Xi$) cascade systems with
upper- and lower-level driving, respectively. Only $\Lambda$- and
upper-level-driven $\Xi$ TLAs exhibit Fano interference-induced suppression of
absorption~\cite{LeeScully2000}. For simplicity, we focus on the $\Lambda$ TLA
to show how the decaying dressed-states formalism yields distinctive
absorption profiles characteristic of EIT  and
ATS~\cite{AnisimovKocharovskaya2008,a:Abi-Salloum2010}, but our approach to
discern EIT from ATS is independent of the choice of TLA so directly
applicable to upper-level-driven $\Xi$-type TLA.

We use a semiclassical description with decay and dephasing rates manually
inserted. The EM response to the probe is proportional to the probe-induced
excited coherence corresponding to the off-diagonal TLA density matrix
element~$\sigma_\text{ab}$. The steady-state solution to linear order of the
probe electric field has all population in~$|b\rangle$ so excited coherence at
the probed transition depends only on dephasing rates~$\Gamma_{\text{ab}}$
and~$\Gamma_{\text{bc}}$: $\sigma _{\text{ab}}=\alpha/[\delta +\Delta
-\text{i}\Gamma_{\text{ab}}-\Omega^2/(\delta -\text{i}\Gamma _{\text{bc}})]$,
with~$\Delta$ the one-photon detuning and~$\alpha$ the probe Rabi
frequency~\cite{AnisimovKocharovskaya2008}. 

Linear absorption~$A\propto\text{Im}(\sigma_{\text{ab}})$, shown in
Figs.~\ref{fig:LS}(b-d), has spectral poles $\delta
_\pm=-\Delta/2+\text{i}(\Gamma _{\text{ab}}+\Gamma
_{\text{bc}})/2\pm[\Omega^2+(\Delta-\text{i}\Gamma_{\text{ab}}+\text{i}\Gamma_{\text{bc}})^2/4]^{1/2}$,
which produce resonant contributions to atomic response,
$A_{\pm}=S_{\pm}/(\delta-\delta_{\pm})$, with strengths
$S_{\pm}=\pm(\delta_{\pm}-\text{i}\Gamma_\text{bc})/(\delta_+-\delta_-)$. These
resonant contributions can be attributed to ``decaying-dressed
states''~\cite{AnisimovKocharovskaya2008} with frequencies and dephasing rates
given by Re$(\delta_{\pm})$ and Im$(\delta_{\pm})$,
respectively. Decaying-dressed states arise from the interaction between
dressed states with eigenenergies $-\Delta/2\pm(\Omega^2+\Delta^2/4)^{1/2}$
and two reservoirs with decay rates $\Gamma_\text{ab}$ and
$\Gamma_\text{bc}$. This interaction is affected by the pump in two ways:
separating dressed-states and exciting the $|a\rangle\leftrightarrow|c\rangle$
transition needed for destructive Fano interference with the
$|a\rangle\leftrightarrow|b\rangle$ reservoir. Unfortunately, the excited
$|a\rangle\leftrightarrow|c\rangle$ transition interacts with the
$|b\rangle\leftrightarrow|c\rangle$ reservoir, which is always positive and
thus negates absorption suppression. Finally, one-photon detuning further
separates dressed states thereby weakening Fano interference. 

Strong Fano interference, hence strong EIT, occurs for resonant driving
($\Delta=0$) where the spectral poles exist in three $\Omega$-regions:
(i)~dressed states share a reservoir
$\Omega\leq\Omega_\text{EIT}\equiv\left(\Gamma_{\text{ab}}-\Gamma_{\text{bc}}\right)/2$,
(ii)~dressed states decay into distinct reservoirs
$\Omega\gg\Gamma_\text{ab}$, and (iii)~intermediate regime where the
dressed-state reservoirs are only partially distinct. In $\Omega$-region (i)
$\text{Re}(\delta _{\pm })=0=\text{Im}(S_{\pm})$ so the absorption profile
comprises two Lorentzians centered at the origin, one broad and positive and
the other narrow and negative:
$A_\text{EIT}=C^2_+/(\gamma_+^2+\delta^2)-C^2_-/(\gamma_-^2+\delta^2)$. Hence,
low-power pump-induced transparency, where Fano interference dominates, has a
transparency window without splitting~\cite{AnisimovKocharovskaya2008}. For
strong-pump $\Omega$-region (ii)
$\delta_\pm\approx\pm\Omega+\text{i}(\Gamma_{\text{ab}}+\Gamma _\text{bc})/2$
and $S_{\pm}\approx1/2$ so  $
	A_\text{ATS} = C^2[1/(\gamma^2+\left(\delta -\delta _0\right)^2)+1/(\gamma^2+\left(\delta +\delta _0\right)^2)]$,
corresponding to the sum of two equal-width Lorentzians shifted from the origin
by $\delta_0\approx\pm \Omega$.

Figures~\ref{fig:LS}(b-d) demonstrate how well these EIT and ATS models fit
calculated absorption profiles, but an objective criterion is needed to
discern the best model or whether the data are inconclusive. 
Akaike's Information Criterion (AIC) identifies the most informative model
based on Kullback-Leibler divergence (relative entropy), which is the average
logarithmic difference between two distributions with respect to the first
distribution. AIC quantifies the information lost when model~$A_i$ with $K_i$
fitting parameters is used to fit actual data: $I_i=-2\log\mathcal{L}_i+2K_i$
for $\mathcal{L}_i$ the maximum likelihood for model~$A_i$ with penalty $2K_i$
for fitting parameters~\cite{BurnhamAnderson2002}. 

\begin{figure}
\includegraphics[width=\columnwidth]{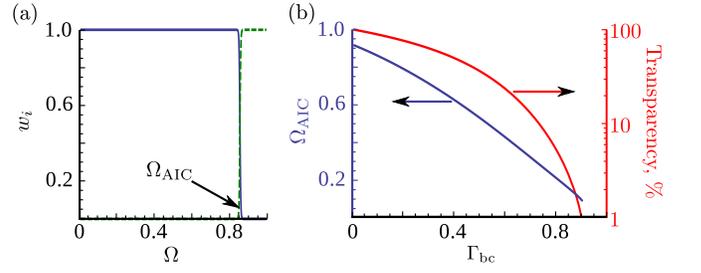}
\caption{(Color online) (a)~Akaike weights vs.\ Rabi frequency for the TLA in
  Fig.~\ref{fig:LS} showing a sharp transition at~$\Omega_\text{AIC}$ from EIT
  model (blue solid) to ATS model (green dashed);
	(b)~Transition boundary~$\Omega_\text{AIC}$ with corresponding
        transparency values vs.\ $\Gamma_\text{bc}$.
	}
\label{fig:AWsGbc}
\end{figure}

We demonstrate AIC-based testing by fitting an absorption data set
$D=\{A(\delta_j);|\delta_j|\le 5\}$, incrementing in steps
$\Delta\delta_j=0.05$, for the TLA in Fig.~\ref{fig:LS}(a) to models
$A_\text{EIT}$ and $A_\text{ATS}$ using the NonlinearModelFit function in
Mathematica\texttrademark, which can calculate AIC.  The relative likelihood
of model~$A_i$ out of~$n$ models is its Akaike weight
$w_i=\text{e}^{-I_i/2}/\sum_{k=1}^n\text{e}^{-I_k/2}$ depicted in
Fig.~\ref{fig:AWsGbc}(a). This figure shows that, based on AIC, the EIT model
explains data with 100\% likelihood for all $\Omega<\Omega_{\rm
  AIC}=0.86$. Figure~\ref{fig:AWsGbc}(b) shows that increasing
$\Gamma_\text{bc}$ reduces the EIT threshold~$\Omega_\text{AIC}$ and guides
devising EIT experiments. 

Testing for EIT is affected by the fact that experiments have additional
complexities such as one-photon detuning or more than three energy levels,
but these complexities do not negate the validity of our test; rather these
complications just make it harder to \emph{pass} the EIT test. Consequently,
one can construct and test more general models that accommodate these extra
features because AIC allows relative testing between any number of models. The
corresponding signatures of Fano interference in generalized models can be
identified thus revealing genuine EIT effects.

A more important issue of working with experimental data sets
$D=\{A(\delta_{j})\}$ is that experiments are noisy so each run produces a
different data set, say $D_\ell$, with many data points measured. In turn, the
Akaike weight reveals the likelihood of describing a data set $D_\ell$ that
becomes binary (0 or 1), hence conclusive, for large data sets as shown in
Fig.~\ref{fig:AWsGbc}(a). Consequently, one will conclusively say after each
run which model pertains, but, because of noise, this conclusion 
could vary from run to run. Intuitively the best model should be picked more
often, however, experimental data are not reported on per run
basis. Experimental data are typically reported as mean values with error bars
representing the confidence interval for the data. Hence we need to adapt the
AIC based testing to the way experimental data are reported.

Akaike's information according to the least-squares analysis is
$I=N\log(\hat{\sigma}^2)+2K$ for
$\hat{\sigma}^2=\sum_{j=1}^{N}\hat{\epsilon}_j^2/N$ and $\hat{\epsilon}^2_j$
the estimated residuals from the fitted model \cite{BurnhamAnderson2002}. 
Technical noise, however, blurs the distinction between models $\{A_i\}$
causing Akaike's information to become
$I=N\log(\hat{\sigma}^2+\hat{\sigma}^2_\text{exp})+2K$ with aforementioned
consequences. Hence, we propose a fitness test for Akaike's information
obtained from reported experimental data.

Our fitness test uses a per-point (mean) AIC contribution $\bar{I}=I/N$ to
calculate a per-point weight for the $i^{\rm th}$ model:
$\bar{w}_i=\exp(-\bar{I}_i/2)/\sum_{k=1}^n\exp(-\bar{I}_k/2)$. These
unnormalized per-point weights $\exp(-\bar{I}_i/2)$ converge to
$1/\sqrt{\hat{\sigma}^2_i}$ for large data sets; in case of noisy data, this
yields equal per-point weights for all models as expected intuitively. 

\begin{figure}
\includegraphics{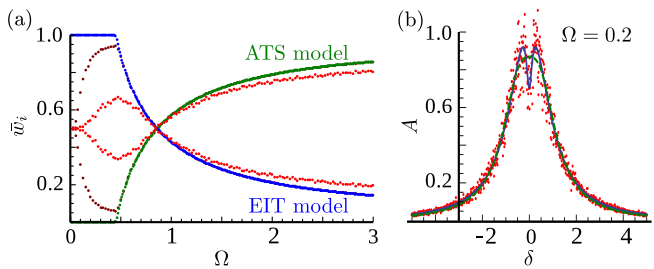}
\caption{(Color online)
	(a)~Per-point weights~$\bar{w}_i$ for the conditions of
        Fig.~\ref{fig:LS} as a function of pump-field Rabi frequency~$\Omega$
        illustrate three distinct regions: $\Omega<\Omega_\text{EIT}=0.45$,
        where the EIT model (blue) dominates unconditionally;
        $0.45<\Omega<0.86$, where the ATS model (green) shows non-zero
        likelihood; $\Omega>0.86$, ATS model dominates. The presence of
        Gaussian noise with standard derivation $\sigma=0.1$ (red dots)
        [$\sigma=0.01$ (burgundy dots)] affects the per-point weights for EIT
        and ATS models leading to the absence of unconditional dominance by
        the EIT model. (b)~In the weak-pump limit and 
        a poor signal-to-noise ratio, both models are equally likely to
        fit data (red dots).}
\label{fig:AWs}
\end{figure}

We simulate a noisy absorption profile by generating data $D_{\ell}$ according to
$\left\langle A(\delta_j)\right\rangle=(1+\xi)A(\delta_j)$
for $\xi$ randomly chosen from the normal distribution 
$\exp\left[-x^2/2 \sigma^2\right]/\sqrt{2 \pi}
\sigma$. Figure~\ref{fig:AWs}(a) shows our per-point weights for generated
data with no noise, small noise and moderate noise for the conditions of
Fig.~\ref{fig:LS}. In no noise case and $\Omega<\Omega_\text{EIT}=0.45$, the ATS
model fails and has per-point weight: $\bar{w}_2=0$; beyond the EIT threshold
$\Omega_\text{EIT}$, the per-point weight for ATS starts to increases with
both models describing the absorption profile equally well at $\Omega_{\rm
  AIC}=0.86$. This agrees with intuition about fitting models, especially a
continuous trade-off between models in the intermediate regime. It is also
intuitive to expect that under noisy conditions and weak pump,
$\Omega^2<\Omega^2_{\sigma}=2\sigma\Gamma_\text{ab}\Gamma_\text{bc}/(1-2\sigma)$,
induced transparency is buried in noise, $1-\text{Im}[\sigma_{\text{ab}}(\delta=0,\Omega)]/\text{Im}[\sigma_{\text{ab}}(\delta=0,\Omega=0)]<2\sigma$,
and both models account for the absorption profile equally
well [see Fig.~\ref{fig:AWs}(b)]. Consequently, at $\Omega=0$ and any amount of
noise, per-point weights are equal to 0.5 and results are
inconclusive. Increasing the pump field, however, favors the EIT model until
it gives way to ATS dominance for pump strength greater than $\Omega_{\rm AIC}$.
Therefore, a convincing EIT demonstration requires suppression of technical
noise to the point that our per-point weights become well separated.

We apply our theory to the recent observation of induced transmission (i.e.\
transparency), reported as EIT, for an open transmission line of a
superconducting circuit with a single flux-type artificial atom (`flux
qubit')~\cite{AbdumalikovTsai2010}. In contrast to TLA system discussed here,
a flux qubit driven/probed by microwave fields, which are polarized and
confined to one dimension, presents a nearly lossless upper-pumped $\Xi$
system. Nevertheless, EIT testing of this observation is straightforward, with absorption being effectively replaced by reflection, since
their analysis shows that transmission coefficient agrees with the electromagnetic response for a TLA: 
$t=1-(\gamma_\text{ab}/2)/[\Gamma_\text{ab}+i\delta+\Omega^2/(\Gamma_\text{bc}+i\delta)]$
with our Rabi frequency~$\Omega$ being half their Rabi
frequency~\cite{AbdumalikovTsai2010}. 

Induced transparency is evident from calculating Re$(t)$ for the probe field
in the presence of the control field. Their system has population relaxation
rate $\gamma_\text{ab}/2\pi=11$~MHz and dephasing rates
$\Gamma_\text{ab}/2\pi=7.2$~MHz and
$\Gamma_\text{bc}=0.96\Gamma_\text{ab}$. Therefore, the transparency window
appears for a control field amplitude of $\Omega/2\pi=6$~MHz, which exceeds
$\Omega_\text{EIT}/2\pi=0.15$~MHz so the experiment operates in a region where
demonstrating Fano interference must be inconclusive.

\begin{figure}
\includegraphics{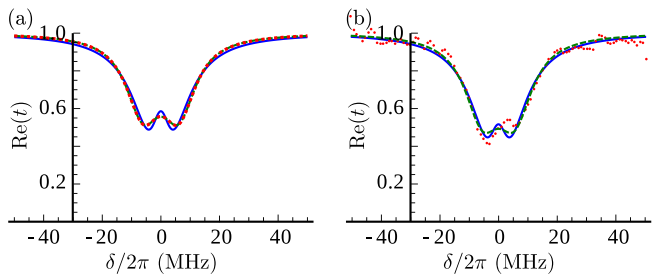}
\caption{(Color online)
	Transmission Re$(t)$ vs.\ two-photon detuning~$\delta$ for
        (a)~theoretical curve (red dots) with parameters taken from
        Ref.~\cite{AbdumalikovTsai2010} and control-field amplitude
        $\Omega=6$~MHz compared to the best-fit $A_\text{EIT}(25.4, 24.2,
        6.36, 6.15)$ (blue solid) and $A_\text{ATS}(4.42, 7.1, 6.1)$ (green
        dashes) and (b)~actual experimental data form
        Ref.~\cite{AbdumalikovTsai2010} (red dots) vs.\ best-fit
        $A_\text{EIT}(11.8, 9.08, 6.77, 5.66)$ (blue solid) and
        $A_\text{ATS}(4.59, 7.29, 5.49)$. 
        }
\label{fig:exper}
\end{figure}

In fact the theoretical transmission curve based on the reported parameters,
shown in Fig.~\ref{fig:exper}(a), is indistinguishable from the best-fit ATS
model and clearly distinct from the EIT model. This is further corroborated by
our per-point weight that yields $\bar{w}_1=0.03$ implying that the
result is far from EIT. Whereas the reported induced transparency suffices for
switching of propagating waves in a superconducting
circuit~\cite{AbdumalikovTsai2010}, our objective test shows conclusively that
they demonstrated ATS and definitely not EIT. 

Due to noise, however, the actual experimental data shown in
Fig.~\ref{fig:exper}(b)~differs from the theoretical prediction discussed
above and shown in Fig.~\ref{fig:exper}(a) so a reported data set does not
conclusively show EIT nor rule it out. That is, optimal choices
of~$A_\text{EIT}$ and~$A_\text{ATS}$ seem to fit the data equally well. Yet,
there is a slight preference for ATS according to our per-point weight
criterion, $\bar{w}_1=0.48$ and $\bar{w}_2=0.52$, in the weak-field limit with
obvious favoring of ATS in the strong-field regime.
 
In conclusion, we propose an objective way to discern ATS vs.\ EIT from
experimental data obtained in systems that demonstrate a smooth transition
from ATS to EIT through three qualitative regions as the strength of the
driving field~$\Omega$ decreases. The Akaike weight provides a rigorous
criterion to ascertain from each data set which of EIT or ATS pertains. We have
introduced a per-point weight that accommodates experimental noise and readily
produces a conclusion of whether EIT or ATS pertain as well as alternative in
case that the experiment is inconclusive. In essence, our test seeks direct evidence of Fano interference, which is manifested as a negative Lorentzian in the
absorption spectrum accompanied by the absence of splitting, that ATS
lacks. Akaike's information criterion, combined with our per-point weights,
allows for testing arbitrarily many models simultaneously. Thus the data can
be tested against a more complicated model, that takes care of additional
levels, one-photon detunings as well as inhomogeneous broadenings, with
greater likelihood of a conclusive result even in the presence of noise, and
we have made clear that the sought-after EIT signature is Fano interference,
which should appear as narrow negative Lorentzians in the data. 

Our test of EIT vs.\ ATS is especially important if operation in the weak-pump
regime is necessary for applications such as sensing so that the data
unambiguously reveal whether the requisite conditions have been met. Nowadays
EIT is becoming demonstrated in a multitude of experimental systems, and a
proper test is needed, which can be employed for any EIT-type experiment. We have provided such a test.

\acknowledgments

BCS acknowledges valuable discussions with A.\ Abdumalikov, Y.\ Nakamura and
P.\ Nussenzweig,
and is especially grateful to A.\ Abdumalikov for providing data to test their
superconducting-circuit induced-transparency experiment. BCS has financial
support from NSERC, \emph{i}CORE and a CIFAR Fellowship. PMA and JPD acknowledge support from ARO, DoE, FQXi, NSF, and the Northrop-Grumman Corporation.


\begin{thebibliography}{23}
\expandafter\ifx\csname natexlab\endcsname\relax\def\natexlab#1{#1}\fi
\expandafter\ifx\csname bibnamefont\endcsname\relax
  \def\bibnamefont#1{#1}\fi
\expandafter\ifx\csname bibfnamefont\endcsname\relax
  \def\bibfnamefont#1{#1}\fi
\expandafter\ifx\csname citenamefont\endcsname\relax
  \def\citenamefont#1{#1}\fi
\expandafter\ifx\csname url\endcsname\relax
  \def\url#1{\texttt{#1}}\fi
\expandafter\ifx\csname urlprefix\endcsname\relax\def\urlprefix{URL }\fi
\providecommand{\bibinfo}[2]{#2}
\providecommand{\eprint}[2][]{\url{#2}}

\bibitem[{\citenamefont{Arimondo}(1996)}]{Arimondo1996}
\bibinfo{author}{\bibfnamefont{E.}~\bibnamefont{Arimondo}},
  \emph{\bibinfo{title}{Coherent population trapping in laser spectroscopy}}
  (\bibinfo{publisher}{Elsevier}, \bibinfo{address}{Amsterdam},
  \bibinfo{year}{1996}), vol.~\bibinfo{volume}{5} of
  \emph{\bibinfo{series}{Prog. Opt.}}, pp. \bibinfo{pages}{257--354}.

\bibitem[{\citenamefont{Kocharovskaya}(1992)}]{Kocharovskaya1992}
\bibinfo{author}{\bibfnamefont{O.}~\bibnamefont{Kocharovskaya}},
  \bibinfo{journal}{Phys. Rep.} \textbf{\bibinfo{volume}{219}},
  \bibinfo{pages}{175} (\bibinfo{year}{1992}).

\bibitem[{\citenamefont{Marangos}(1998)}]{Marangos1998}
\bibinfo{author}{\bibfnamefont{J.~P.} \bibnamefont{Marangos}},
  \bibinfo{journal}{J. Mod. Opt.} \textbf{\bibinfo{volume}{45}},
  \bibinfo{pages}{471} (\bibinfo{year}{1998}).

\bibitem[{\citenamefont{Xu et~al.}(2008)\citenamefont{Xu, Sun, Berman, Steel,
  Bracker, Gammon, and Sham}}]{Xu2008}
\bibinfo{author}{\bibfnamefont{X.}~\bibnamefont{Xu}},
  \bibinfo{author}{\bibfnamefont{B.}~\bibnamefont{Sun}},
  \bibinfo{author}{\bibfnamefont{P.~R.} \bibnamefont{Berman}},
  \bibinfo{author}{\bibfnamefont{D.~G.} \bibnamefont{Steel}},
  \bibinfo{author}{\bibfnamefont{A.~S.} \bibnamefont{Bracker}},
  \bibinfo{author}{\bibfnamefont{D.}~\bibnamefont{Gammon}}, \bibnamefont{and}
  \bibinfo{author}{\bibfnamefont{L.~J.} \bibnamefont{Sham}},
  \bibinfo{journal}{Nat. Phys.} \textbf{\bibinfo{volume}{4}},
  \bibinfo{pages}{692} (\bibinfo{year}{2008}).

\bibitem[{\citenamefont{Liu et~al.}(2009)\citenamefont{Liu, Langguth, Weiss,
  Kastel, Fleischhauer, Pfau, and Giessen}}]{Liu2009}
\bibinfo{author}{\bibfnamefont{N.}~\bibnamefont{Liu}},
  \bibinfo{author}{\bibfnamefont{L.}~\bibnamefont{Langguth}},
  \bibinfo{author}{\bibfnamefont{T.}~\bibnamefont{Weiss}},
  \bibinfo{author}{\bibfnamefont{J.}~\bibnamefont{Kastel}},
  \bibinfo{author}{\bibfnamefont{M.}~\bibnamefont{Fleischhauer}},
  \bibinfo{author}{\bibfnamefont{T.}~\bibnamefont{Pfau}}, \bibnamefont{and}
  \bibinfo{author}{\bibfnamefont{H.}~\bibnamefont{Giessen}},
  \bibinfo{journal}{Nat. Mater.} \textbf{\bibinfo{volume}{8}},
  \bibinfo{pages}{758} (\bibinfo{year}{2009}).

\bibitem[{\citenamefont{Kelly et~al.}(2010)\citenamefont{Kelly, Dutton,
  Schlafer, Mookerji, Ohki, Kline, and Pappas}}]{Kelly2010}
\bibinfo{author}{\bibfnamefont{W.~R.} \bibnamefont{Kelly}},
  \bibinfo{author}{\bibfnamefont{Z.}~\bibnamefont{Dutton}},
  \bibinfo{author}{\bibfnamefont{J.}~\bibnamefont{Schlafer}},
  \bibinfo{author}{\bibfnamefont{B.}~\bibnamefont{Mookerji}},
  \bibinfo{author}{\bibfnamefont{T.~A.} \bibnamefont{Ohki}},
  \bibinfo{author}{\bibfnamefont{J.~S.} \bibnamefont{Kline}}, \bibnamefont{and}
  \bibinfo{author}{\bibfnamefont{D.~P.} \bibnamefont{Pappas}},
  \bibinfo{journal}{Phys. Rev. Lett.} \textbf{\bibinfo{volume}{104}},
  \bibinfo{pages}{163601} (\bibinfo{year}{2010}).

\bibitem[{\citenamefont{Papasimakis et~al.}(2008)\citenamefont{Papasimakis,
  Fedotov, Zheludev, and Prosvirnin}}]{Papasimakis2008}
\bibinfo{author}{\bibfnamefont{N.}~\bibnamefont{Papasimakis}},
  \bibinfo{author}{\bibfnamefont{V.~A.} \bibnamefont{Fedotov}},
  \bibinfo{author}{\bibfnamefont{N.~I.} \bibnamefont{Zheludev}},
  \bibnamefont{and} \bibinfo{author}{\bibfnamefont{S.~L.}
  \bibnamefont{Prosvirnin}}, \bibinfo{journal}{Phys. Rev. Lett.}
  \textbf{\bibinfo{volume}{101}}, \bibinfo{pages}{253903}
  (\bibinfo{year}{2008}).

\bibitem[{\citenamefont{Zhang et~al.}(2008)\citenamefont{Zhang, Genov, Wang,
  Liu, and Zhang}}]{Zhang2008}
\bibinfo{author}{\bibfnamefont{S.}~\bibnamefont{Zhang}},
  \bibinfo{author}{\bibfnamefont{D.~A.} \bibnamefont{Genov}},
  \bibinfo{author}{\bibfnamefont{Y.}~\bibnamefont{Wang}},
  \bibinfo{author}{\bibfnamefont{M.}~\bibnamefont{Liu}}, \bibnamefont{and}
  \bibinfo{author}{\bibfnamefont{X.}~\bibnamefont{Zhang}},
  \bibinfo{journal}{Phys. Rev. Lett.} \textbf{\bibinfo{volume}{101}},
  \bibinfo{pages}{047401} (\bibinfo{year}{2008}).

\bibitem[{\citenamefont{Safavi-Naeini et~al.}(2011)\citenamefont{Safavi-Naeini,
  Alegre, Chan, Eichenfield, Winger, Lin, Hill, Chang, and Painter}}]{SMC+11}
\bibinfo{author}{\bibfnamefont{A.~H.} \bibnamefont{Safavi-Naeini}},
  \bibinfo{author}{\bibfnamefont{T.~P.~M.} \bibnamefont{Alegre}},
  \bibinfo{author}{\bibfnamefont{J.}~\bibnamefont{Chan}},
  \bibinfo{author}{\bibfnamefont{M.}~\bibnamefont{Eichenfield}},
  \bibinfo{author}{\bibfnamefont{M.}~\bibnamefont{Winger}},
  \bibinfo{author}{\bibfnamefont{Q.}~\bibnamefont{Lin}},
  \bibinfo{author}{\bibfnamefont{J.~T.} \bibnamefont{Hill}},
  \bibinfo{author}{\bibfnamefont{D.~E.} \bibnamefont{Chang}}, \bibnamefont{and}
  \bibinfo{author}{\bibfnamefont{O.}~\bibnamefont{Painter}},
  \bibinfo{journal}{Nature} \textbf{\bibinfo{volume}{472}}, \bibinfo{pages}{69}
  (\bibinfo{year}{2011}).

\bibitem[{\citenamefont{Lamb and Retherford}(1951)}]{a:LambRetherford1951}
\bibinfo{author}{\bibfnamefont{W.~E.} \bibnamefont{Lamb}} \bibnamefont{and}
  \bibinfo{author}{\bibfnamefont{R.~C.} \bibnamefont{Retherford}},
  \bibinfo{journal}{Phys. Rev.} \textbf{\bibinfo{volume}{81}},
  \bibinfo{pages}{222} (\bibinfo{year}{1951}).

\bibitem[{\citenamefont{Alzar et~al.}(2002)\citenamefont{Alzar, Martinez, and
  Nussenzveig}}]{a:Nussenzveig2002}
\bibinfo{author}{\bibfnamefont{C.~L.~G.} \bibnamefont{Alzar}},
  \bibinfo{author}{\bibfnamefont{M.~A.~G.} \bibnamefont{Martinez}},
  \bibnamefont{and}
  \bibinfo{author}{\bibfnamefont{P.}~\bibnamefont{Nussenzveig}},
  \bibinfo{journal}{Am. J. Phys.} \textbf{\bibinfo{volume}{70}},
  \bibinfo{pages}{37} (\bibinfo{year}{2002}).

\bibitem[{\citenamefont{Fleischhauer et~al.}(2005)\citenamefont{Fleischhauer,
  Imamo\u{g}lu, and Marangos}}]{FleischhauerMarangos2005}
\bibinfo{author}{\bibfnamefont{M.}~\bibnamefont{Fleischhauer}},
  \bibinfo{author}{\bibfnamefont{A.}~\bibnamefont{Imamo\u{g}lu}},
  \bibnamefont{and} \bibinfo{author}{\bibfnamefont{J.~P.}
  \bibnamefont{Marangos}}, \bibinfo{journal}{Rev. Mod. Phys.}
  \textbf{\bibinfo{volume}{77}}, \bibinfo{pages}{633} (\bibinfo{year}{2005}).

\bibitem[{\citenamefont{Hau et~al.}(1999)\citenamefont{Hau, Harris, Dutton, and
  Behroozi}}]{Hau1999}
\bibinfo{author}{\bibfnamefont{L.~V.} \bibnamefont{Hau}},
  \bibinfo{author}{\bibfnamefont{S.~E.} \bibnamefont{Harris}},
  \bibinfo{author}{\bibfnamefont{Z.}~\bibnamefont{Dutton}}, \bibnamefont{and}
  \bibinfo{author}{\bibfnamefont{C.~H.} \bibnamefont{Behroozi}},
  \bibinfo{journal}{Nature} \textbf{\bibinfo{volume}{397}},
  \bibinfo{pages}{594} (\bibinfo{year}{1999}).

\bibitem[{\citenamefont{Phillips et~al.}(2001)\citenamefont{Phillips,
  Fleischhauer, Mair, Walsworth, and Lukin}}]{Philips2001}
\bibinfo{author}{\bibfnamefont{D.~F.} \bibnamefont{Phillips}},
  \bibinfo{author}{\bibfnamefont{A.}~\bibnamefont{Fleischhauer}},
  \bibinfo{author}{\bibfnamefont{A.}~\bibnamefont{Mair}},
  \bibinfo{author}{\bibfnamefont{R.~L.} \bibnamefont{Walsworth}},
  \bibnamefont{and} \bibinfo{author}{\bibfnamefont{M.~D.} \bibnamefont{Lukin}},
  \bibinfo{journal}{Phys. Rev. Lett.} \textbf{\bibinfo{volume}{86}},
  \bibinfo{pages}{783} (\bibinfo{year}{2001}).

\bibitem[{\citenamefont{Fano}(1961)}]{Fano1961}
\bibinfo{author}{\bibfnamefont{U.}~\bibnamefont{Fano}}, \bibinfo{journal}{Phys.
  Rev.} \textbf{\bibinfo{volume}{124}}, \bibinfo{pages}{1866}
  (\bibinfo{year}{1961}).

\bibitem[{\citenamefont{Autler and Townes}(1955)}]{AutlerTownes1955}
\bibinfo{author}{\bibfnamefont{S.~H.} \bibnamefont{Autler}} \bibnamefont{and}
  \bibinfo{author}{\bibfnamefont{C.~H.} \bibnamefont{Townes}},
  \bibinfo{journal}{Phys. Rev.} \textbf{\bibinfo{volume}{100}},
  \bibinfo{pages}{703} (\bibinfo{year}{1955}).

\bibitem[{\citenamefont{Burnham and Anderson}(2002)}]{BurnhamAnderson2002}
\bibinfo{author}{\bibfnamefont{K.~P.} \bibnamefont{Burnham}} \bibnamefont{and}
  \bibinfo{author}{\bibfnamefont{D.~R.} \bibnamefont{Anderson}},
  \emph{\bibinfo{title}{Model Selection and Multimodel Inference}}
  (\bibinfo{publisher}{Springer-Verlag}, \bibinfo{address}{New York},
  \bibinfo{year}{2002}), \bibinfo{edition}{2nd} ed.

\bibitem[{\citenamefont{Hao et~al.}(2008)\citenamefont{Hao, Sonnefraud, Dorpe,
  Maier, Halas, and Nordlander}}]{Hao2008}
\bibinfo{author}{\bibfnamefont{F.}~\bibnamefont{Hao}},
  \bibinfo{author}{\bibfnamefont{Y.}~\bibnamefont{Sonnefraud}},
  \bibinfo{author}{\bibfnamefont{P.~V.} \bibnamefont{Dorpe}},
  \bibinfo{author}{\bibfnamefont{S.~A.} \bibnamefont{Maier}},
  \bibinfo{author}{\bibfnamefont{N.~J.} \bibnamefont{Halas}}, \bibnamefont{and}
  \bibinfo{author}{\bibfnamefont{P.}~\bibnamefont{Nordlander}},
  \bibinfo{journal}{Nano Lett.} \textbf{\bibinfo{volume}{8}},
  \bibinfo{pages}{3983} (\bibinfo{year}{2008}).

\bibitem[{\citenamefont{Imamo\u{g}lu and Harris}(1989)}]{ImamogluHarris89}
\bibinfo{author}{\bibfnamefont{A.}~\bibnamefont{Imamo\u{g}lu}}
  \bibnamefont{and} \bibinfo{author}{\bibfnamefont{S.~E.}
  \bibnamefont{Harris}}, \bibinfo{journal}{Opt. Lett.}
  \textbf{\bibinfo{volume}{14}}, \bibinfo{pages}{1344} (\bibinfo{year}{1989}).

\bibitem[{\citenamefont{Lee et~al.}(2000)\citenamefont{Lee, Rostovtsev, and
  Scully}}]{LeeScully2000}
\bibinfo{author}{\bibfnamefont{H.}~\bibnamefont{Lee}},
  \bibinfo{author}{\bibfnamefont{Y.}~\bibnamefont{Rostovtsev}},
  \bibnamefont{and} \bibinfo{author}{\bibfnamefont{M.~O.}
  \bibnamefont{Scully}}, \bibinfo{journal}{Phys. Rev. A}
  \textbf{\bibinfo{volume}{62}}, \bibinfo{pages}{063804}
  (\bibinfo{year}{2000}).

\bibitem[{\citenamefont{Anisimov and
  Kocharovskaya}(2008)}]{AnisimovKocharovskaya2008}
\bibinfo{author}{\bibfnamefont{P.}~\bibnamefont{Anisimov}} \bibnamefont{and}
  \bibinfo{author}{\bibfnamefont{O.}~\bibnamefont{Kocharovskaya}},
  \bibinfo{journal}{J. Mod. Opt.} \textbf{\bibinfo{volume}{55}},
  \bibinfo{pages}{3159 } (\bibinfo{year}{2008}).

\bibitem[{\citenamefont{Abi-Salloum}(2010)}]{a:Abi-Salloum2010}
\bibinfo{author}{\bibfnamefont{T.~Y.} \bibnamefont{Abi-Salloum}},
  \bibinfo{journal}{Phys. Rev. A} \textbf{\bibinfo{volume}{81}},
  \bibinfo{pages}{053836} (\bibinfo{year}{2010}).

\bibitem[{\citenamefont{Abdumalikov et~al.}(2010)\citenamefont{Abdumalikov,
  Astafiev, Zagoskin, Pashkin, Nakamura, and Tsai}}]{AbdumalikovTsai2010}
\bibinfo{author}{\bibfnamefont{A.~A.} \bibnamefont{Abdumalikov}},
  \bibinfo{author}{\bibfnamefont{O.}~\bibnamefont{Astafiev}},
  \bibinfo{author}{\bibfnamefont{A.~M.} \bibnamefont{Zagoskin}},
  \bibinfo{author}{\bibfnamefont{Y.~A.} \bibnamefont{Pashkin}},
  \bibinfo{author}{\bibfnamefont{Y.}~\bibnamefont{Nakamura}}, \bibnamefont{and}
  \bibinfo{author}{\bibfnamefont{J.~S.} \bibnamefont{Tsai}},
  \bibinfo{journal}{Phys. Rev. Lett.} \textbf{\bibinfo{volume}{104}},
  \bibinfo{pages}{193601} (\bibinfo{year}{2010}).

\end{thebibliography}

\end{document}